\documentclass{aip-cp}

\usepackage{lineno}
\usepackage[numbers]{natbib}
\usepackage[mathletters]{ucs}   
\usepackage[utf8x]{inputenc}
\usepackage{rotating}
\usepackage{booktabs}
\usepackage{graphicx}
\usepackage{xcolor}
\usepackage{xspace}
\usepackage{overpic}
\usepackage{siunitx}
\DeclareSIUnit\pe{p.e.}
\DeclareSIUnit\ev{eV}
\DeclareSIUnit\kev{\kilo\eV}
\DeclareSIUnit\mev{\mega\eV}
\DeclareSIUnit\gev{\giga\eV}
\DeclareSIUnit\tev{\tera\eV}
\DeclareSIUnit\erg{erg}
\DeclareSIUnit\parsec{pc}
\DeclareSIUnit\year{yr}
\DeclareSIUnit{\msun}{\mbox{$M_{\odot}$}}

\newcommand{\ct}[1]{CT\,#1\xspace}
\newcommand{\hess}{H.E.S.S.\xspace}

\newcommand{\psrls}{PSR B1259-63/LS 2883\xspace}
\newcommand{\he}{HE\xspace}

\newcommand{\vhe}{VHE\xspace}
\newcommand{\vhes}{VHEs\xspace}
\newcommand{\newbinary}{HESS J1832-093\xspace}

\begin{document}

\definecolor{hessred}{RGB}{211,27,23}

\title{Observations of Binary Systems with the \hess Telescopes}

\author[aff2]{P. Bordas}
\author[aff3]{G. Dubus}
\author[aff2]{P. Eger}
\author[aff5]{J.-P. Ernenwein}
\author[aff6]{H. Laffon}
\author[aff7]{C. Mariaud}
\author[aff1]{T. Murach\corref{cor1}}
\author[aff7]{M. de Naurois}
\author[aff9]{C. Romoli}
\author[aff10]{F. Sch\"ussler}
\author[aff2]{R. Zanin}
\author{for the \hess Collaboration}

\affil[aff1]{Institut f\"ur Physik, Humboldt-Universit\"at zu Berlin, Newtonstr. 15, D 12489 Berlin, Germany}
\affil[aff2]{Max-Planck-Institut f\"ur Kernphysik, P.O. Box 103980, D 69029 Heidelberg, Germany}
\affil[aff3]{Univ. Grenoble Alpes, IPAG,  F-38000 Grenoble, France \protect\\ CNRS, IPAG, F-38000 Grenoble, France}
\affil[aff5]{Aix Marseille Universit\'e, CNRS/IN2P3, CPPM UMR 7346,  13288 Marseille, France}
\affil[aff6]{Universit\'e Bordeaux, CNRS/IN2P3, Centre d'\'Etudes Nucl\'eaires de Bordeaux Gradignan, 33175 Gradignan, France}
\affil[aff7]{Laboratoire Leprince-Ringuet, Ecole Polytechnique, CNRS/IN2P3, F-91128 Palaiseau, France}
\affil[aff9]{Dublin Institute for Advanced Studies, 31 Fitzwilliam Place, Dublin 2, Ireland}
\affil[aff10]{DSM/Irfu, CEA Saclay, F-91191 Gif-Sur-Yvette Cedex, France}
\corresp[cor1]{Corresponding author: murach@physik.hu-berlin.de}

\maketitle

\begin{abstract}
Observations of binary systems obtained recently with the High Energy Stereoscopic System (H.E.S.S) of Cherenkov telescopes are reported. The outcomes of a
detailed observation campaign on PSR B1259-63 during its periastron passage in 2014 will be presented. This system was observed for the first time with
H.E.S.S. II, providing spectra and light curves down to 200 GeV, which will be compared with observations conducted during previous periastron passages and
with results from an analysis of contemporaneously taken Fermi-LAT data. Also long-term observations of LS 5039 with H.E.S.S in phase I and phase II are reported.
This source was monitored at very high energies (\vhes) in a period of time spanning more than ten years. Its spectral energy distribution measured with
\hess~II extends down to \SI{120}{\gev}. Spectral results from the Fermi-LAT observations are shown as well, and the compatibility with \hess results in the overlapping
energy range is discussed. The identification of the new gamma-ray binary candidate \newbinary will also be presented. Furthermore, the search for \vhe emission from the
microquasars GRS 1915+105, Circinus X-1 and V4641 Sgr based on data from H.E.S.S. observations conducted contemporaneously with the \emph{RXTE} satellite
experiment will be reported. These data provide constraints on the integral gamma-ray flux at different X-ray states of the three sources.
\end{abstract}

\section{Introduction}
A majority of \SI{70}{\percent} of all stars in our galaxy are part of binary systems \cite{mathieu_habil}. Depending on the orbital parameters, the nature
of the individual objects and properties of the emission, these systems can be grouped into several categories. The following sections will focus on results
from observations of gamma-ray binaries and microquasars (MQs) in the \vhe (very high energy, with photon energies above $\sim\SI{100}{\gev}$) gamma-ray domain.
The former are binary systems whose non-thermal radiative output is dominated by gamma rays, i.e. photons with energies above $\sim\SI{1}{\mev}$. They consist of
a compact object, either a neutron star or a stellar-mass black hole, and a massive O- or B-type star. Only seven of such systems are known to date.
MQs are binary systems which emit X-rays and- exhibit extended radio emission. High-energy (\he, $\SI{100}{\mev}<E<\SI{100}{\gev}$) gamma-ray emission has
been reported from at least two such systems, Cyg X-1 and Cyg X-3. However, a detection of \vhe gamma-ray emission at the 5σ level is still lacking.

Both gamma-ray binaries and microquasars exhibit flux variability and sometimes, especially in case of the former, periodicity at all wavelengths. Details
depend on the parameters of the systems, for example on the orbital periods which can last from a few days up to a few years, influencing the location of the
interaction zone and the energy and particle densities in this area. Hence these systems provide a large variety of observables, making them interesting
targets for observations and subjects of theoretical models.

The results presented here are derived from observations performed with the High Energy Stereoscopic System (\hess) telescopes. In phase~I of the \hess experiment,
regular observations were conducted with four telescopes with \SI{107}{\meter\squared} reflectors each arranged in a square with a side length of \SI{120}{\meter}
started in 2004. In 2012, a much larger fifth telescope with a reflector area of \SI{596}{\meter\squared} called \ct{5} was placed into the centre of the
array, marking the beginning of phase II of the \hess experiment. In these proceedings, results from observations conducted with both the \hess-I and
the \hess-II configurations are reported. For more information about the experimental setup, see e.g. \citet{hesscrab}.

In the following sections, results from observations of \psrls around the 2014 periastron in are presented. Furthermore, results from
long-term observations of LS 5039 are reported. Lastly, the discovery of a gamma-ray binary candidate is discussed and results from observations of three
microquasars are presented.

\section{Observations of \psrls}
\psrls is a binary system comprising a massive O9.5Ve star with a circumstellar disk and a pulsar with a rotational period of \SI{48}{\milli\second} and a
spin-down luminosity of \SI{8e35}{\erg\per\second}. The orbit of the pulsar around the central star is very eccentric, with an eccentricity of $e=0.87$, and
has a period of \SI{3.4}{\year}. The system is located at a distance of \SI{2.3}{\kilo\parsec}. It is the only gamma-ray binary system for which the nature of
the compact object is known \cite{psr_discovery,psrb1259_23yrs_timing}.

At \vhes, \psrls was discovered with the \hess telescopes around the 2004 periastron \cite{psr_hess_discovery}. Since then, observations with the \hess
telescopes were conducted around all periastron passages since then \cite{psrb1259_hess_2007,psrb1259_hess_2010}. In case of the 2014 periastron passage, data equivalent to \SI{60}{\hour} of
live time were collected with the five-telescope \hess-II array. Both the pre- and post-periastron phases were covered as well as, for the first time,
the time of periastron itself. Gamma rays from this system were detected with a statistical significance of $>30σ$.

\begin{figure}[ht]
  \centerline{\begin{overpic}[width=0.5\textwidth]{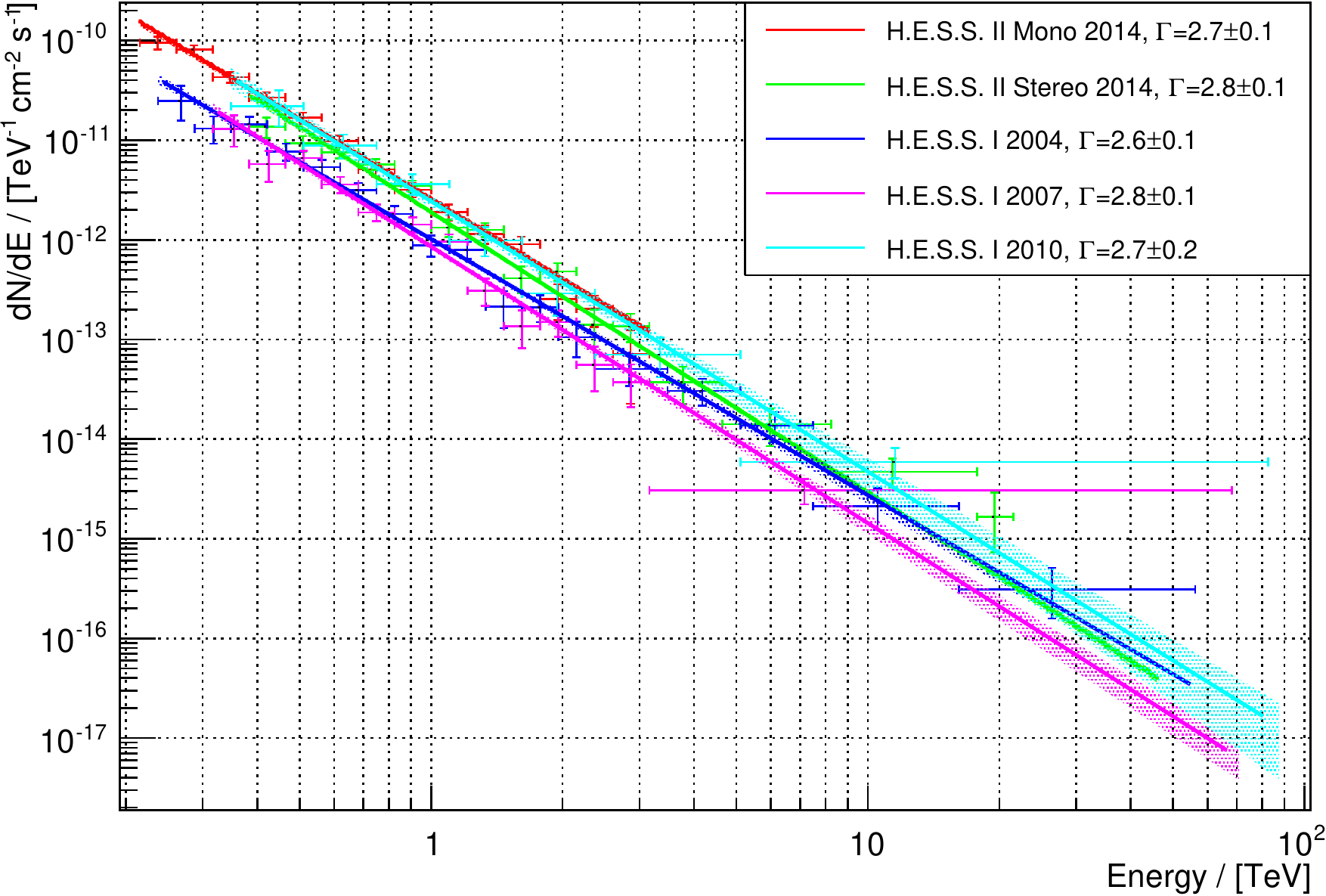}
      \put(12,8){\textbf{\textcolor{hessred}{\hess Preliminary}}}
    \end{overpic}}
  \caption{Differential photon spectra obtained from monoscopic and stereoscopic analyses of the 2014 data set. Also spectra resulting from re-analyses
           of archival data from previous periastra are shown. Spectral indices are given in the legend.}
  \label{fig:spectra}
\end{figure}

The differential photon spectrum obtained from a monoscopic analysis of the 2014 data performed with the MonoReco algorithm \cite{monoreco} is shown
in Fig.~\ref{fig:spectra} together with results from an extension of the Hillas reconstruction presented in \citet{hesscrab} enabling analyses of data
recorded with the \hess II array.
Spectral indices as given in the figure are compatible with each other. Also the flux normalisations at \SI{1}{\tev} are compatible
within systematic uncertainties assumed to be \SI{20}{\percent} of the flux. For comparison, also spectra from re-analyses of archival data from
observations around previous periastra are shown. These spectra are obtained from analyses with the ImPACT algorithm \cite{impact}. While all spectral
indices are compatible with each other within errors, the flux normalisation differs from one year to the other. This is expected, as different parts
of the orbit were sampled each year. It should be noted that the safe energy threshold of the
re-analyses of archival data are signifiantly lower than the published thresholds due to refined analysis techniques. The monoscopic safe energy
threshold is expected to be lowered by several tens of GeV in the future, since the current value of approximately \SI{200}{\gev} is chosen arbitrarily
to guarantee robust results.

\begin{figure}[ht]
  \begin{overpic}[width=0.49\textwidth]{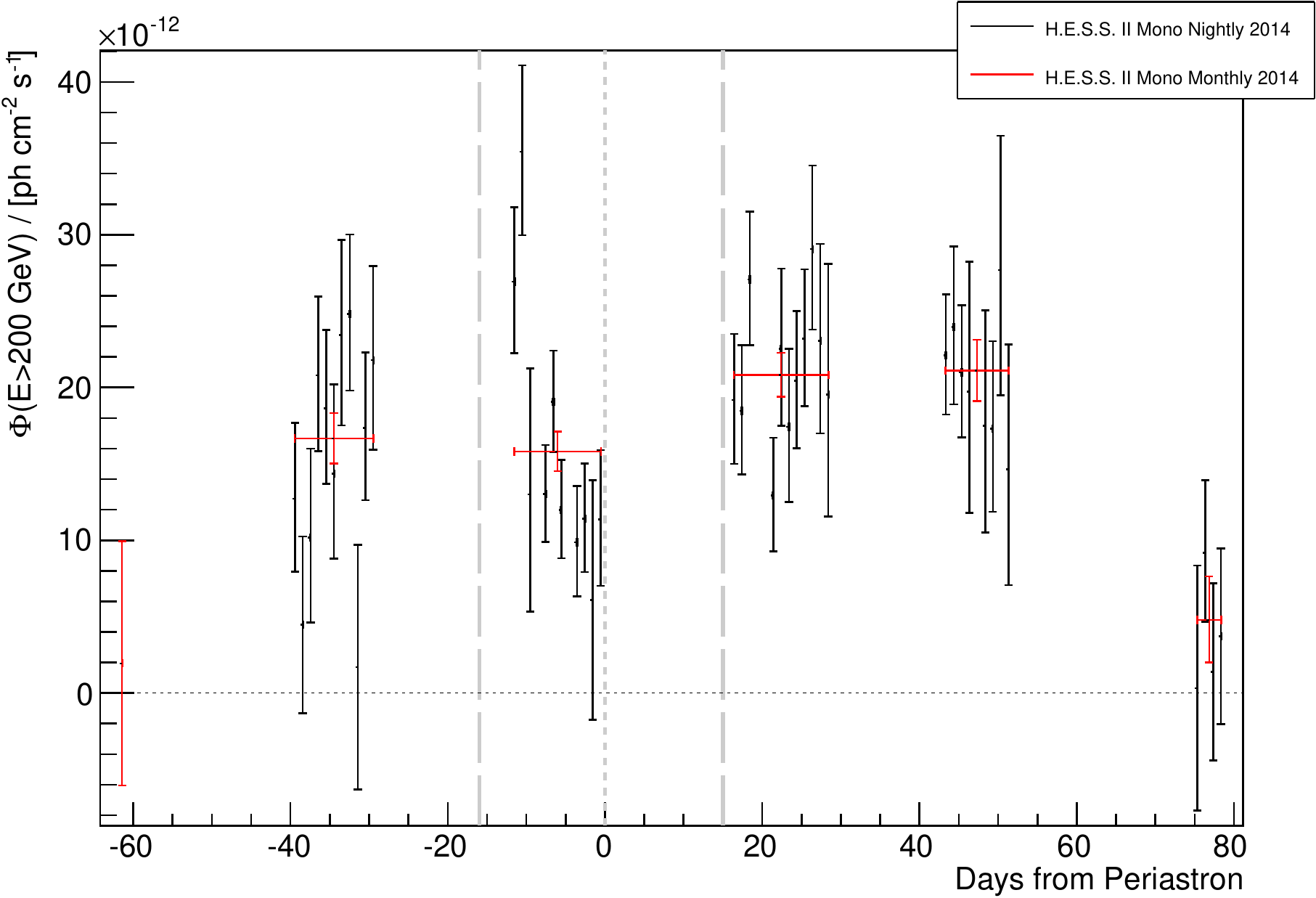}
    \put(10,59){\textbf{\textcolor{hessred}{\hess Preliminary}}}
  \end{overpic}\quad
  \begin{overpic}[width=0.49\textwidth]{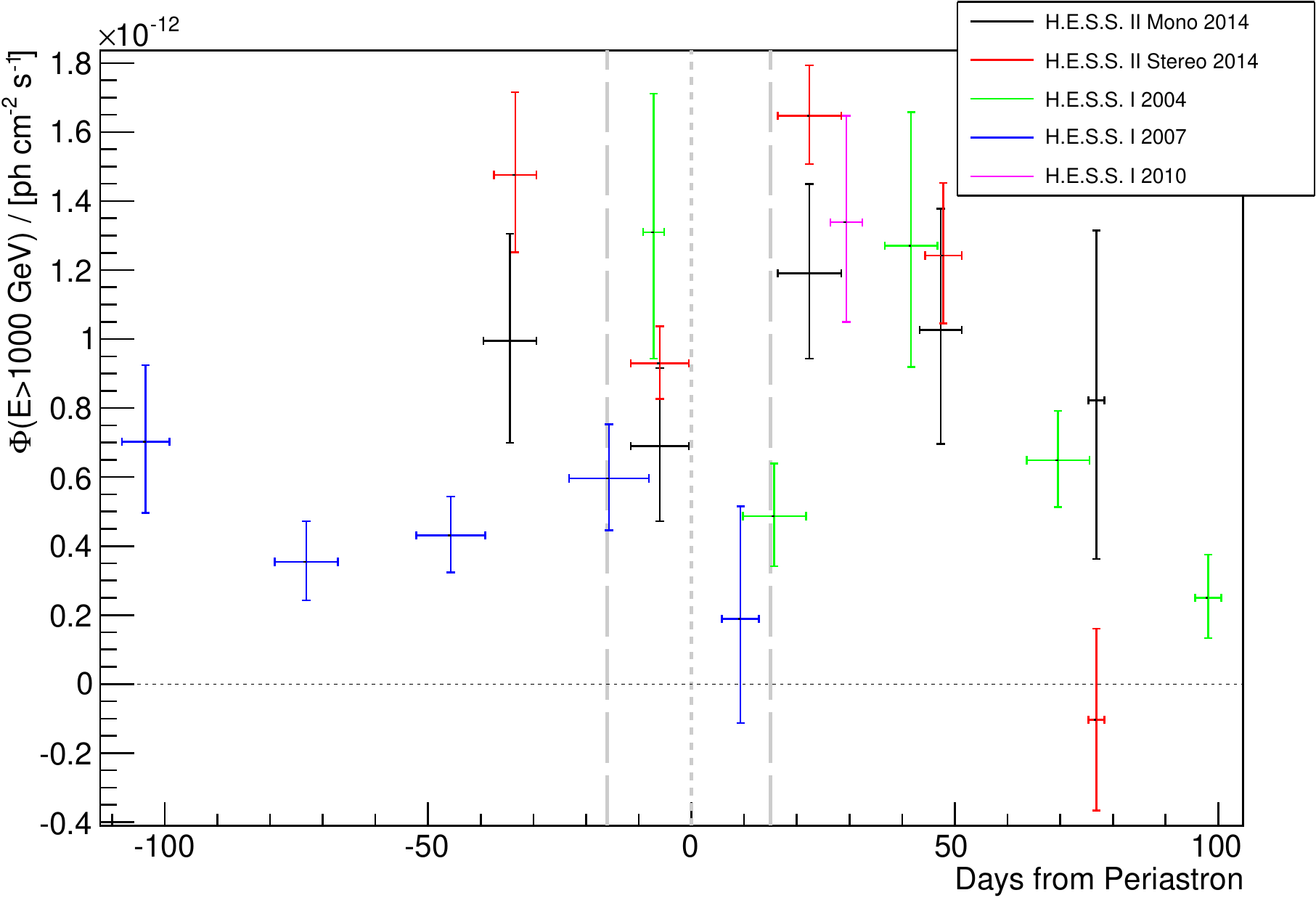}
    \put(10,59){\textbf{\textcolor{hessred}{\hess Preliminary}}}
  \end{overpic}
  \caption{Light curves above \SI{200}{\gev} (\textit{left}) and \SI{1}{\tev} (\textit{right}) with night-wise and monthly binnings. The times of the
           disk crossings and of the periastron are marked by grey dashed and dotted lines, respectively.}
  \label{fig:psr_LCs}
\end{figure}

Fluxes above \SI{200}{\gev} are shown in the left panel of Fig.~\ref{fig:psr_LCs} with a night-wise and a monthly binning. In the night-wise light curve,
several interesting features can be identified. A rising flux well before the first disk crossing is observed, with maximum fluxes compatible to the
flux level observed around the actual disk crossings. After the first disk crossing, the flux decreases towards periastron. Finally a high flux, again
with fluxes similar to the fluxes observed during the disk crossings, is seen at the time of the GeV flare occurring between \SI{32}{\day} and approximately
\SI{60}{\day} after periastron. For comparison also light curves obtained from all years are shown on the right panel of Fig.~\ref{fig:psr_LCs} for
energies above \SI{1}{\tev}. The general features previously discussed remain the same. It should be noted, though, that the high flux before the first
disk crossing is followed by a low flux at the time of the nominal disk crossing. This behaviour is still under investigation. Furthermore it can be seen
that also above \SI{1}{\tev} the flux is high at the central part of the GeV flare, which may constrain current theoretical models of the origin of the
periodic \he flare.

\begin{figure}[ht]
  \centerline{\begin{overpic}[width=0.6\textwidth]{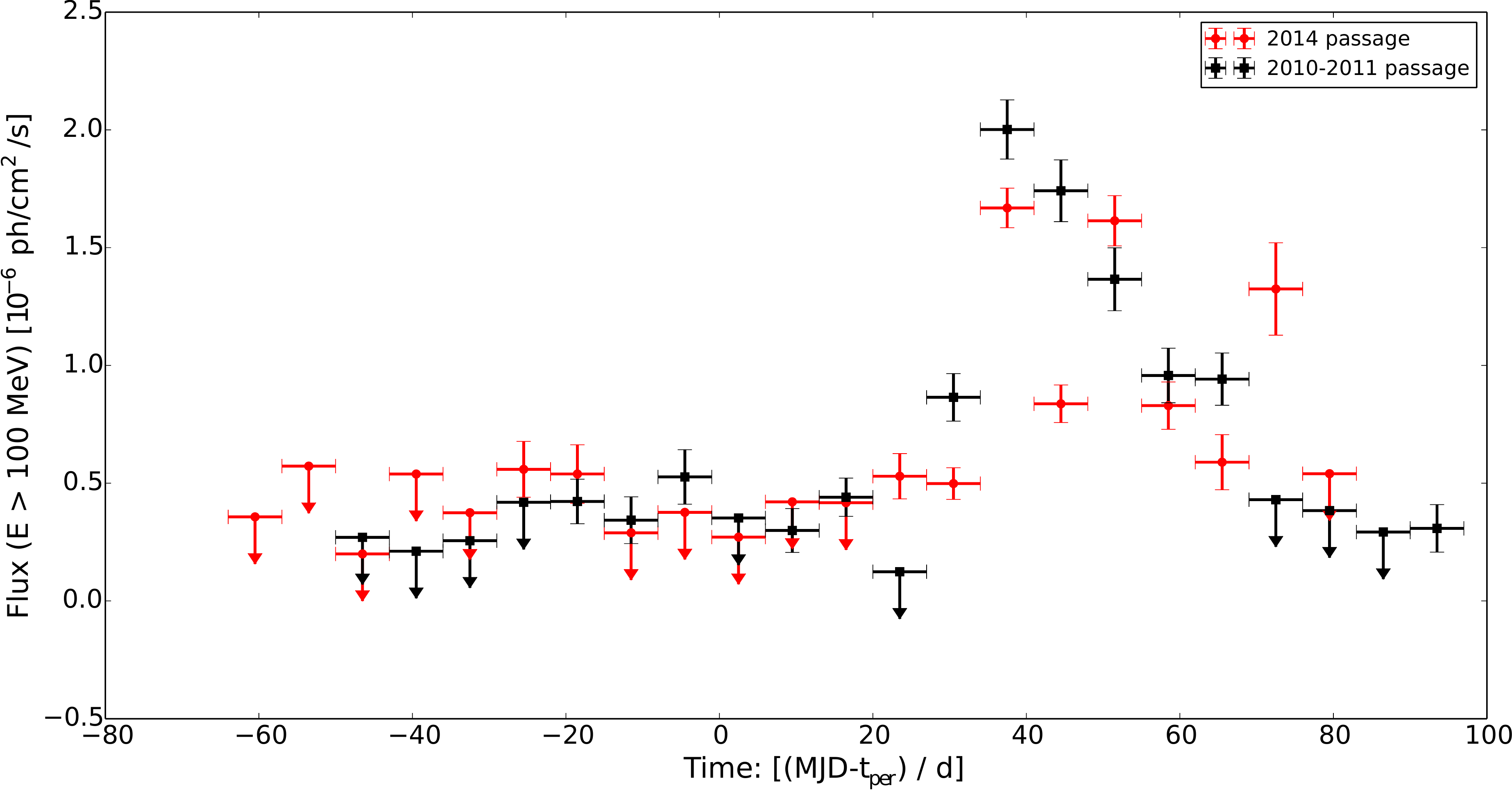}
      \put(10,47){\textbf{\textcolor{hessred}{\hess Preliminary}}}
    \end{overpic}}
  \caption{Fermi-LAT light curve with a weekly binning obtained from a re-analyses of the 2010 and 2014 data sets based on the \emph{Pass 8} release.
           Upper limits are indicated by arrows.}
  \label{fig:fermi_lc}
\end{figure}

Data from measurements with Fermi-LAT corresponding to the release version \emph{Pass 8} were analysed as well. The obtained light curve is shown in
Fig.~\ref{fig:fermi_lc}.
Apart from the fact that the prominent flares observed approximately \SI{30}{\day} after periastron in both 2010 and 2014 can be reproduced it should be
noted that, for the first time, several flux points with a significance of at least 3σ were obtained for the times around the disk crossings in both
years. There is no indication of a high flux state at the time of the high flux state at \vhes before the first disk crossing.

\section{Observations of LS 5039}
The gamma-ray binary LS 5039 consists of an O6.5Ve star with a mass of \SI{23}{\msun} and a compact object of unknown nature with a mass of
\SI{3.7}{\msun}. These two objects are in a very close orbit with a period of only \SI{3.9}{\day}. In this section, results from ten years of \hess
observations are presented. The data set comprises \SI{104}{\hour} and \SI{19}{\hour} of live time recorded with the \hess array in phase~I
and phase~II, respectively.

\begin{figure}[ht]
  \includegraphics[width=0.49\textwidth]{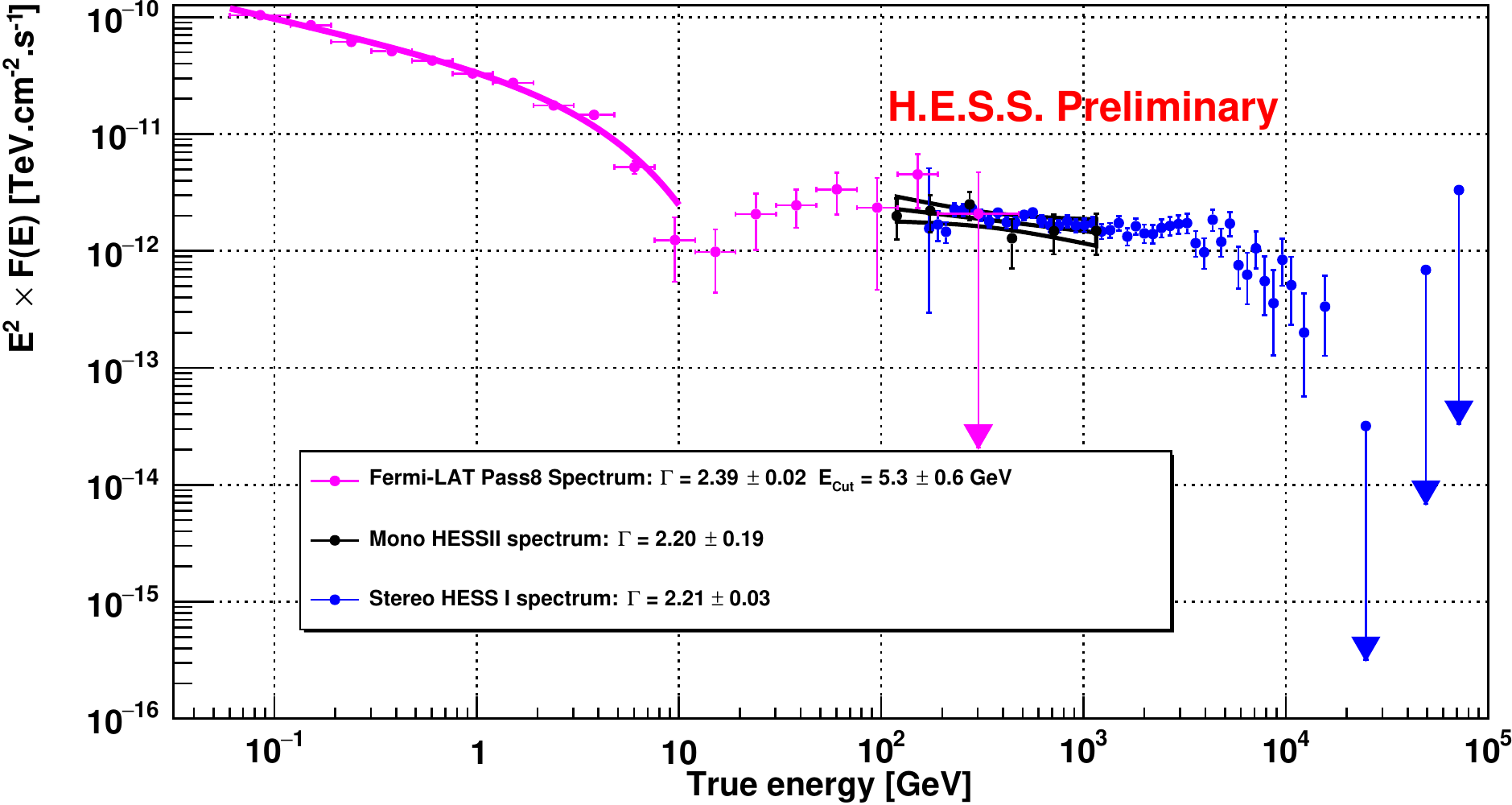}\quad
  \includegraphics[width=0.49\textwidth]{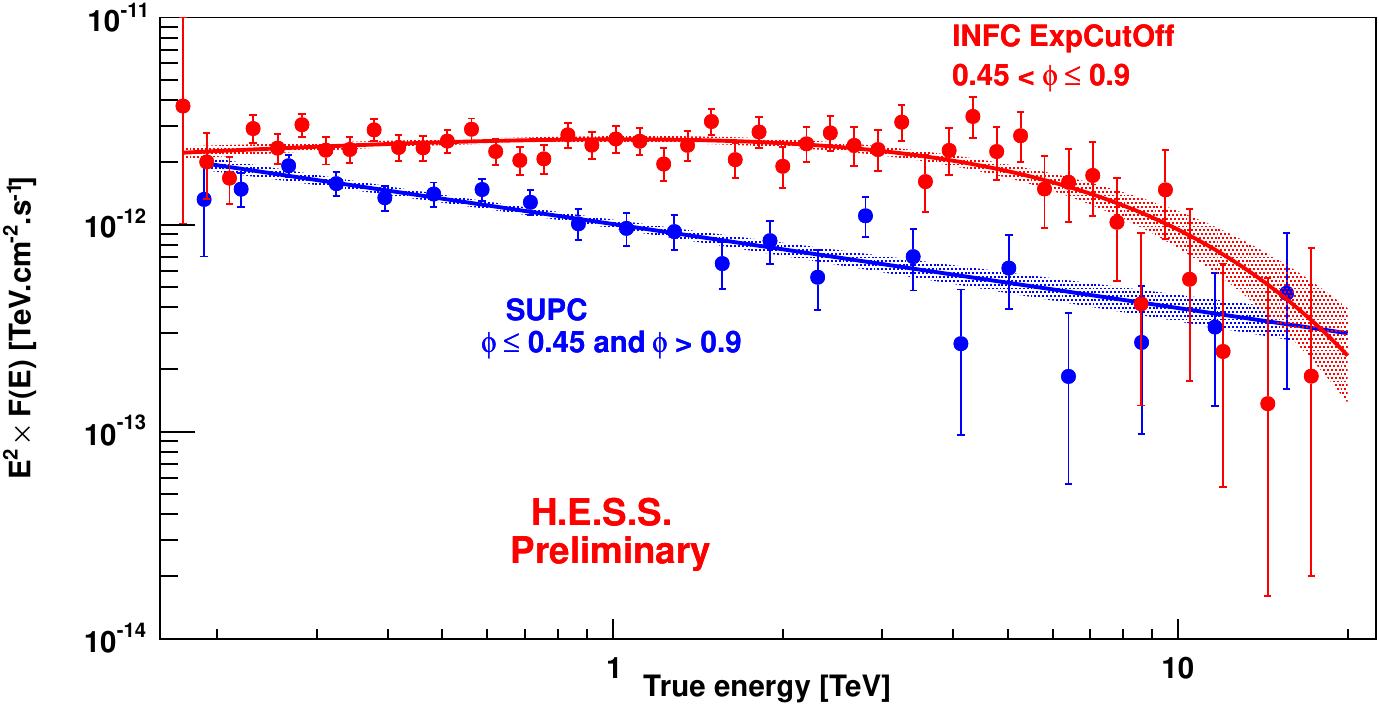}
  \caption{\textit{Left}: SEDs obtained from monoscopic and a stereoscopic analyses of the \hess-II and \hess-I data sets, respectively. Results of
           fits with power-law functions are given in the inset. Also an SED obtained from a re-analysis of Fermi-LAT data is shown. \textit{Right}: SEDs
           resulting from \hess-I analyses for parts of the orbit corresponding to the inferior or superior conjunction. The corresponding orbital
           phase ranges are given for reference. Fit results are given in the main text.}
  \label{fig:ls5039_spectra}
\end{figure}

Analyses of all available data yield the spectral energy distributions (SEDs) shown on the left side of Fig.~\ref{fig:ls5039_spectra}. Both an
SED resulting from a monoscopic analysis of the data taken with the \hess~II array and an SED obtained from a stereoscopic analysis of \hess~I data
are shown. The analyses were performed with the \emph{Model Analysis} \cite{modelpp}.
In case of the monoscopic analysis, the energy threshold is approximately \SI{120}{\gev}.
Together with the \hess curves, also an SED obtained from a re-analysis of Fermi-LAT data in the \emph{Pass 8} release is shown. For the first time
the energy ranges of LS 5039 SEDs resulting from the two experiments significantly overlap. The data points in this region are compatible with each
other within statistical uncertainties.

On the right side of Fig.~\ref{fig:ls5039_spectra}, two other SEDs resulting from an analysis of the \hess-I data set are shown. The two SEDs
correspond to either the inferior or the superior conjunction phase of the orbit. While the latter SED is well-fit by a power law, the former one
exhibits a curvature and is therefore fit with a power law with an exponential cut-off at approximately \SI{6.6+-1.6}{\tev}. This cut-off energy
is approximately 1000 times the cut-off energy of \SI{5.3+-0.6}{\gev} observed in the \he gamma-ray spectrum shown on the left side of the figure. The
indices of the spectra are $Γ=\num{2.41+-0.06}$ and $Γ=\num{1.84+-0.07}$ for the superior and inferior conjunctions, respectively. At
\SI{1}{\tev}, the normalisations of the differential photon fluxes are \SI{1.01+-0.05e-12}{ph\per\tev\per\centi\meter\squared\per\second} and
\SI{2.58+-0.1e-12}{ph\per\tev\per\centi\meter\squared\per\second}.

For more information, the reader is referred to \citet{ls5039_proceeding}.

\section{Observations of \newbinary}
\newbinary was detected at \vhes during the galactic plane scan performed with the \hess experiment \cite{j1832_discovery}. The observed gamma-ray
excess in the direction of object source is compatible with the excess from a point-like source of \vhe gamma rays. The differential photon spectrum
is well described by a power law with an index of \num{2.6+-0.3}. Several scenarios about the nature of this object are discussed in the literature.
Here the tentative identification as one of the seven aforementioned gamma-ray binaries is presented.

Recent measurements with the Chandra satellite experiment reveal that the apparent angular distance of the star 2MASS J18324516−0921545 and the X-ray
source XMMU J183245−0921539 is only \ang{;;0.3}, suggesting an association of these two objects due to the low chance coincidence of \SI{0.3}{\percent}.
Also the high column density inferred from a spectral fit of the X-ray data, which is approximately ten times higher than the total galactic value,
supports the hypothesis that the objects form a binary system.
The fact that \newbinary is compatible with a point-like source at \vhes disfavours the identification as a pulsar wind nebula or a supernova remnant,
since such objects appear, in general, as extended sources. Furthermore, the flux observed at X-ray energies is variable, with a factor of approximately
six between the low and high flux states, which is similar to the value obtained from the gamma-ray binary HESS J0632+057. Another similarity to the
latter system is the fact that no significant \he gamma-ray emission could be detected with current-generation instruments. Despite the lack of such
emission, the spectral energy distribution is dominated by the \vhe emission, which is a necessary requirement for the identification as a gamma-ray
binary.

Based on the above arguments, \newbinary has been catalogued as a gamma-ray binary system. For further information and the discussion of associations with
different object classes, the reader is referred to \citet{j1832_binary}.



\section{Observations of Microquasars}
\begin{figure}[ht]
  \centerline{\includegraphics[width=0.5\textwidth]{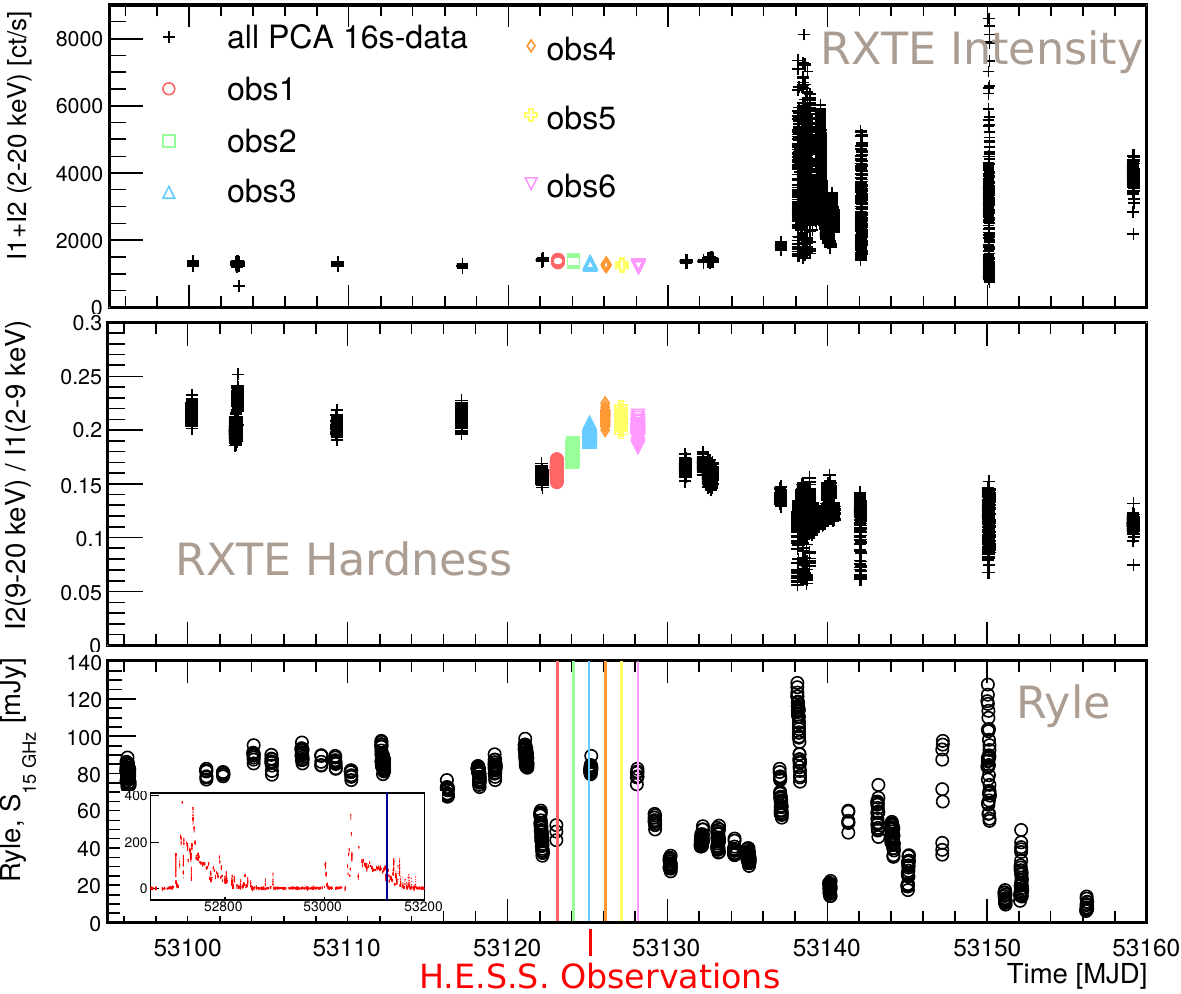}}
  \caption{\textit{Top panel}: Integrated flux between \SI{2}{\kev} and \SI{20}{\kev} as a function of time as measured by \emph{RXTE}. \textit{Central panel}:
           Hardness, defined as the ratio of the fluxes in the higher and lower energy band of the X-ray flux as indicated on the ordinate as a
           function of time. \textit{Bottom panel}: Radio flux at \SI{15}{\giga\hertz} as a function of time as measured by the Ryle experiment.
           Colours denote the times during which \hess observations were performed. Adapted from \citet{MQs}.}
  \label{fig:mq_xray_flux}
\end{figure}

As mentioned earlier, \he gamma rays have been detected from MQs already. \vhe gamma rays are predicted to be emitted by MQs during transient outbursts
by various theoretical models, but have not been detected yet. In this section, results from \hess observations of three MQs conducted contemporaneously
to observations with the \emph{RXTE}
satellite experiment are presented. All observations were performed with the \hess~I array in 2004, aiming to cover times of expected broadband flaring
events. One target of the observations was GRS 1915+105. Observations of this object were performed after a flux decrease in the radio band at frequencies
of approximately \SI{15}{\giga\hertz} between April 28 and May 3, 2004 as shown in Fig.~\ref{fig:mq_xray_flux}.
Cir X-1 was observed around the time of periastron between June 18 and June 20. V4641 Sgr was observed after a rapid increase of the intensity was
measured in the radio, optical and X-ray domains between July 7 and July 9. None of the observations yielded the detection of a significant flux in the
\vhe gamma-ray regime. Thus flux upper limits at the \SI{99}{\percent} confidence level were calculated after an analysis of the data with the Hillas
reconstruction algorithm. The results are displayed in Tab.~\ref{tab:mq_uls}.

For more information, the reader is referred to \citet{MQs}.
\begin{table}[ht]
  \caption{Flux upper limits at the \SI{99}{\percent} confidence level. Limits were calculated assuming a power-law shape of the spectrum with an index
           of 2.6 or 2.0 for the std and hard cut configurations defined in \citet{hesscrab}, respectively. The live time $T_L$, the energy threshold
           $E_{\textrm{thr}}$ and the flux $I$ are listed in the table. Adapted from \citet{MQs}.}
  \label{tab:mq_uls}
  \tabcolsep7pt\begin{tabular}{l|lccc}
    \toprule
      & \tch{1}{c}{b}{Cuts}  & \tch{1}{c}{b}{$T_L / [\si{\hour}]$}  & \tch{1}{c}{b}{$E_{\textrm{thr}} / [\si{GeV}]$}  & \tch{1}{c}{b}{$I(>E_{\textrm{thr}}) / [\si{ph\per\centi\meter\squared\per\second}]$}   \\
    \midrule
    \textbf{GRS 1915+105} & std & 6.9 & 562 & $<\num{7.338e-13}$ \\
       & hard & 6.9 & 1101 & $<\num{1.059e-13}$ \\
    \midrule
    \textbf{Circinus X-1} & std & 5.4 & 562 & $<\num{1.172e-12}$ \\
       & hard & 5.4 & 1101 & $<\num{4.155e-13}$ \\
    \midrule
    \textbf{V4641 Sgr} & std & 1.8 & 237 & $<\num{4.477e-12}$ \\
       & hard & 1.8 & 422 & $<\num{4.795e-13}$ \\
    \bottomrule
  \end{tabular}
\end{table}

\section{Summary}
In these proceedings, results from observations of binary systems with the \hess telescopes have been presented. The gamma-ray binary \psrls was observed around
the 2014 periastron, exhibiting a high flux state at the time of the \si{\gev} flare. A high flux state before first disk crossing is currently under
investigation. There are no clear signatures for inter-orbital variation of the spectral index or the flux level at \vhes. The spectrum could be extended down
to an energy of \SI{200}{\gev}.
In case of LS 5039, results from ten years of observations with both the \hess-I and \hess-II arrays have been presented. Significant spectral variation
is observed between orbital phases corresponding to the inferior and superior conjunctions, respectively. The phase-averaged SEDs obtained from analyses
of both \hess-I and \hess-II data agree well with each other and also with the phase-averaged SED resulting from a re-analysis of Fermi-LAT data.
The identification of the \vhe gamma-ray emitter \newbinary as a gamma-ray binary is discussed as well. This identification is mostly based on features
of the spectral energy distribution, flux variability and the source morphology. Lastly, also observations of MQs are presented. These observations
yield no significant detection of \vhe gamma ray photons, but competitive upper limits for the emission of such photons from microquasars have been computed.



\section{Acknowledgements}
{\small The support of the Namibian authorities and of the University of Namibia in facilitating the construction and operation of H.E.S.S. is gratefully acknowledged, as is the support by the German Ministry for Education and Research (BMBF), the Max Planck Society, the German Research Foundation (DFG), the French Ministry for Research, the CNRS-IN2P3 and the Astroparticle Interdisciplinary Programme of the CNRS, the U.K. Science and Technology Facilities Council (STFC), the IPNP of the Charles University, the Czech Science Foundation, the Polish Ministry of Science and Higher Education, the South African Department of Science and Technology and National Research Foundation, the University of Namibia, the Innsbruck University, the Austrian Science Fund (FWF), and the Austrian Federal Ministry for Science, Research and Economy, and by the University of Adelaide and the Australian Research Council. We appreciate the excellent work of the technical support staff in Berlin, Durham, Hamburg, Heidelberg, Palaiseau, Paris, Saclay, and in Namibia in the construction and operation of the equipment. This work benefited from services provided by the H.E.S.S. Virtual Organisation, supported by the national resource providers of the EGI Federation.}


\nocite{*}
\bibliographystyle{aipnum-cp}%
\bibliography{sample}%

\end{document}